\documentclass[twocolumn,prl,showpacs,superscriptaddress]{revtex4}

\usepackage{graphicx}%
\usepackage{dcolumn}
\usepackage{amsmath}

\usepackage{color}

\makeatletter
\def\btt#1{\texttt{\@backslashchar#1}}%
\DeclareRobustCommand\bblash{\btt{\@backslashchar}}%
\makeatother

\begin{document}
\preprint{HEP/123-qed}

\title[Short Title]{Terahertz radiation by ultrafast spontaneous polarization modulation in multiferroic BiFeO$_{\bf 3}$ thin films}

\author{Kouhei Takahashi}
\affiliation{Institute of Laser Engineering, Osaka University, 2-6 Yamadaoka, Suita, Osaka 565-0876, Japan}

\author{Noriaki Kida}
\affiliation{Spin Superstructure Project (SSS), ERATO, Japan Science and Technology Agency (JST), c/o National Institute of Advanced Industrial Science and Technology (AIST), AIST Tsukuba Central 4, 1-1-1 Higashi, Tsukuba, Ibaraki 305-8562, Japan}

\author{Masayoshi Tonouchi}
\affiliation{Institute of Laser Engineering, Osaka University, 2-6 Yamadaoka, Suita, Osaka 565-0876, Japan}

\date{\today}

\begin{abstract}
Terahertz (THz) radiation has been observed from multiferroic BiFeO$_3$ thin films via ultrafast modulation of spontaneous polarization upon carrier excitation with illumination of femtosecond laser pulses. The radiated THz pulses from BiFeO$_3$ thin films were clarified to directly reflect the spontaneous polarization state, giving rise to a memory effect in a unique style and enabling THz radiation even at zero-bias electric field. On the basis of our findings, we demonstrate potential approaches to ferroelectric nonvolatile random access memory with nondestructive readability and ferroelectric domain imaging microscopy using THz radiation as a sensitive probe.
\end{abstract}

\pacs{77.84.-s, 77.80.Fm, 42.65.Re}

\maketitle

Ferroelectric oxides offer a wide range of functionalities demonstrating their potential for various electronic and photonic applications such as capacitors for data storage devices, piezoelectric actuators, and nonlinear optical devices for frequency conversion of laser lights \cite{1}. Among them, BiFeO$_3$ with a perovskite structure has recently attracted much attention for the presence of multiferroism at room temperature with simultaneous ferroelectricity and antiferromagnetism, as well as for the extremely large spontaneous polarization $P_{\rm s}$ observed in thin film form \cite{2,3}. BiFeO$_3$ is a charge-transfer insulator exhibiting an energy gap of about 2.5 eV \cite{4}, which is quite small in contrast with the ordinary ferroelectrics whose energy gap generally lies in the ultraviolet region. Such distinct feature allows carrier excitation in BiFeO$_3$ with commercially available femtosecond laser pulses, and hence, enables us to develop ferroelectric ultrafast optoelectronic devices as widely demonstrated in semiconductors.

In this Letter, we report a novel terahertz (THz) radiation characteristic of multiferroic BiFeO$_3$ thin films via exciting charge carriers upon illumination of femtosecond laser pulses. Based on the bias electric field dependence of the THz amplitude, we show that the radiated THz pulse directly reflects the $P_{\rm s}$ state indicating that THz radiation results from the ultrafast modulation of $P_{\rm s}$ of the BiFeO$_3$ thin films. This phenomenon can be applied universally to other ferroelectrics and provides new approaches to various potential applications such as readout in nonvolatile random access memories and ferroelectric domain imaging systems.

Ultrafast laser pulses enable us to produce a wide range of unique phenomena in condensed matter. THz radiation via exciting charge carriers upon illumination of ultrafast laser pulses is a typical example, which provides considerable interest owing to a variety of applications such as imaging for biomedical diagnosis and security, time-domain spectroscopy for material characterization, and a tool to evaluate the ultrafast dynamics of photoexcited charge carriers \cite{5,6}. Intensive studies have therefore been carried out on THz radiation from photoconductive switches fabricated on various materials. The mechanism of THz radiation from photoconductive switch is dominated by the dynamical motion of the free carriers created by laser illumination, in which the transient current surge developed within subpicosecond time scale upon carrier excitation gives rise to electromagnetic radiation at THz frequencies. Through this approach, voltage-biased semiconductors have shown the best radiation efficiency among others demonstrating their potential as a general THz radiation source \cite{6}. On the other hand, transition metal oxides with perovskite structures have presented peculiar THz radiation characteristics originating from their strong electron correlations as exemplified by the cases of high-transition-temperature cuprate superconductors \cite{7,8,9} and colossal magnetoresistive manganites \cite{10,11,12,13}. The present work on multiferroic  BiFeO$_3$ thin films provides an alternate route to THz radiation based on this photoconductive approach.

200 nm-thick BiFeO$_3$ thin films with (00$l$) orientation were grown on (LaAlO$_3$)$_{0.3}$(Sr$_2$AlTaO$_6$)$_{0.7}$(001) substrates, abbreviated as LSAT, by pulsed laser deposition technique with a KrF excimer laser at a growth condition of 800$^\circ$C for substrate temperature and 60 Pa for oxygen pressure. The ferroelectricity and the antiferromagnetism of the obtained BiFeO$_3$ thin films at room temperature were confirmed by a ferroelectric tester and superconducting quantum interface device magnetometer, respectively.

A dipole-type photoconductive switch consisting of a pair of Au electrodes was fabricated on the film by a conventional photolithography and sputtering method. The Au electrode of the photoconductive switch consists of a pair of 30 $\mu$m-wide strip lines separated by 20 $\mu$m with a dipole gap of 10 $\mu$m.

\begin{figure}[bt]
\includegraphics[width=0.38\textwidth]{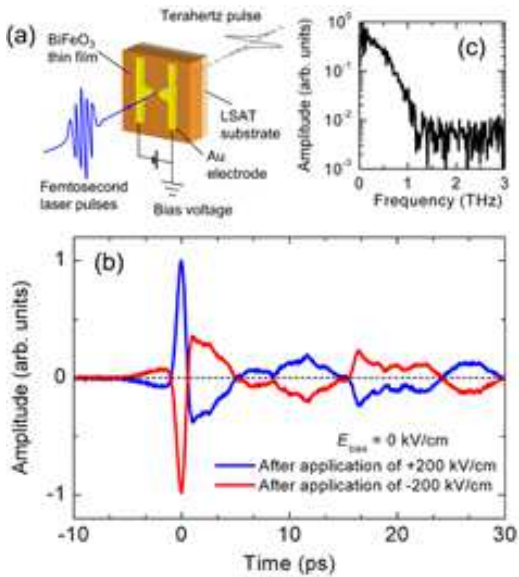}
\caption{(color) (a) Schematic of terahertz (THz) radiation from a BiFeO$_3$ photoconductive switch. (b) Time-domain waveform of the THz pulse radiated from a BiFeO$_3$ photoconductive switch measured at zero-bias electric field after applying a bias electric field $E_{\rm bias}$ of $\pm$200 kV/cm. The dashed line represents the zero-level line shown for clarity. (c) Fourier transformed amplitude spectrum of the THz waveforms in (b).}
\label{fig1}
\end{figure}

In Fig. \ref{fig1}(a), we present the schematic illustration of THz radiation from a photoconductive switch fabricated on a BiFeO$_3$ thin film. Generation of a single-cycle THz pulse propagating into free space was achieved by carrier excitation upon illumination of femtosecond laser pulses with a center wavelength $\lambda$ of 400 nm. We used the second harmonic of a mode-locked Ti:Sapphire laser with a repetition rate of 82 MHz, $\lambda$ of 800 nm, and pulse duration of 100 fs, which was generated through a BaB$_2$O$_4$ crystal. The laser power at $\lambda$ of 400 nm was fixed at 5.5 mW and focused to a spot diameter of 20 $\mu$m. We employed a conventional photoconductive sampling technique using a bow-tie-type low-temperature-grown GaAs photoconductive switch as a detector \cite{6}. All measurements were performed at room temperature.

Figure \ref{fig1}(b) shows a typical example of two time-domain THz waveforms radiated from the BiFeO$_3$ photoconductive switch in the absence of an external bias electric field $E_{\rm bias}$, measured after once applying $E_{\rm bias}$ of $\pm$200 kV/cm. The two THz waveforms measured under the same condition but with different initial treatment, are nearly identical, both consisting of a single-cycle pulse centered at 0 ps with a pulse width of about 0.84 ps, except that they have a reversed phase by $\pi$ with one another. Thus, they show an identical Fourier transformed amplitude spectrum exhibiting a frequency component extending up to 1 THz [Fig. \ref{fig1}(c)]. These features where THz radiation is observed even at zero-bias electric field and its phase depends on the polarity of the initially applied $E_{\rm bias}$, provide a ferroelectric peculiarity that the history of the applied $E_{\rm bias}$ is memorized and the remnant polarization of BiFeO$_3$ substitutes for $E_{\rm bias}$, which is essential for THz radiation. This is in contrast to the case of photoconductive switches fabricated on nonferroelectric materials, where it requires constant application of $E_{\rm bias}$ to radiate a THz pulse \cite{6}.

THz radiation can also be achieved from transparent noncentrosymmetric materials through the second-order nonlinear optical effect when they satisfy a severe phase-matching condition under the incidence of high-intensity laser pulses \cite{6}. However, this common effect observed in ferroelectrics does not contribute to the THz radiation presented here. We examined the response of BiFeO$_3$ thin films to the illumination of laser pulses at $\lambda$ of 800 nm, which show negligible absorption in BiFeO$_3$ \cite{4}. Despite the high laser power of 180 mW illuminated onto the film (this is about twenty times larger than the case at $\lambda$ of 400 nm), THz radiation was not observed in this situation, indicating that THz radiation presented in Fig \ref{fig1}(b) is definitely triggered by the photoexcited charge carriers and not by the nonlinear optical effect, or the optically induced polarization transient as reported in ferroelectric LiNbO$_3$ \cite{14} and LiTaO$_3$ \cite{15}.

\begin{figure}[bt]
\includegraphics[width=0.38\textwidth]{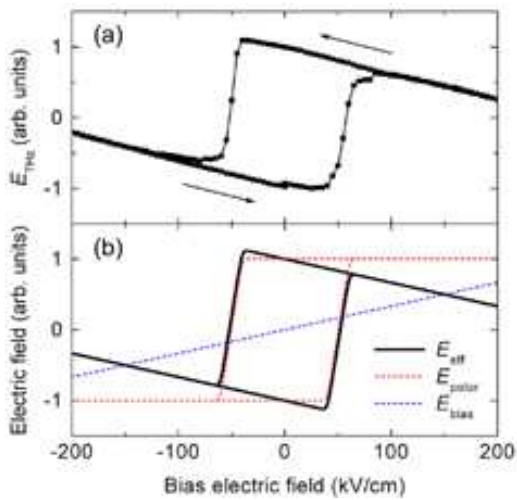}
\caption{(color) Electric polarization of BiFeO$_3$ thin film probing with terahertz (THz) radiation. (a) The main peak amplitude of the radiated THz pulse $E_{\rm THz}$ as a function of the applied bias electric field. The measurement was performed by fixing the laser spot onto the dipole gap. The arrows indicate the sequential directions. (b) An ideal model of the macroscopic electric field induced to a pair of electrodes in typical ferroelectrics for understanding the tilted hysteresis loop in (a). $E_{\rm eff}$, $E_{\rm polar}$, and $E_{\rm bias}$ represents the effective electric field induced between the electrodes, electric field derived by electric polarization, and applied bias electric field, respectively. $E_{\rm eff}$ is expressed by $E_{\rm polar}-E_{\rm bias}$.}
\label{fig2}
\end{figure}

Further showing that THz radiation reflects the $P_{\rm s}$ state of the BiFeO$_3$ thin film, we measured the main peak amplitude of the THz pulse $E_{\rm THz}$ as a function of $E_{\rm bias}$ [Fig. \ref{fig2}(a)]. In the photoconductive switches reported so far \cite{6}, which are all nonferroelectric, $E_{\rm THz}$ generally has linear relationship with the applied $E_{\rm bias}$ reversing its phase by $\pi$ when the polarity of $E_{\rm bias}$ is changed, without showing any hysteresis. However, in the case of BiFeO$_3$, we observed a clear hysteresis loop [Fig. \ref{fig2}(a)], which looks familiar with the common electric polarization hysteresis loop observed in ferroelectrics except for the slight tilted shape where the intensity of $E_{\rm THz}$ is suppressed at high $E_{\rm bias}$. We explain such hysteresis loop by considering the macroscopic electric field biased to a pair of electrodes in typical ferroelectrics [Fig. \ref{fig2}(b)]. The effective macroscopic electric field $E_{\rm eff}$ induced to a pair of electrodes is ideally expressed by $E_{\rm eff}=E_{\rm polar}-E_{\rm bias}$, where $E_{\rm polar}$ represents the electric field derived from electric polarization (= $P$/$\epsilon_0$; $P$: electric polarization and $\epsilon_0$: electric constant). Due to the screening of $E_{\rm polar}$ by $E_{\rm bias}$, $E_{\rm eff}$ is suppressed at high $E_{\rm bias}$ and shows a tilted hysteresis loop in analogy with the $E_{\rm THz}$ loop observed in Fig. \ref{fig2}(a). These features show the direct relationship of THz radiation with $P_{\rm s}$, which emerges as a result of ultrafast $P_{\rm s}$ modulation introduced by the mobile photoexcited charge carriers. We assume that the ferroelectric ordering is not disarranged upon photoexcitation here because only a small amount of photons is injected into the film compared to the total number of the unit cell covered by the laser spot, i.e., the number of injected photons per cubic centimeter is considered to be of the order of 10$^{17}$ per pulse, whereas the number of the perovskite unit cells per cubic centimeter extends up to the order of 10$^{22}$.

\begin{figure}[tt]
\includegraphics[width=0.38\textwidth]{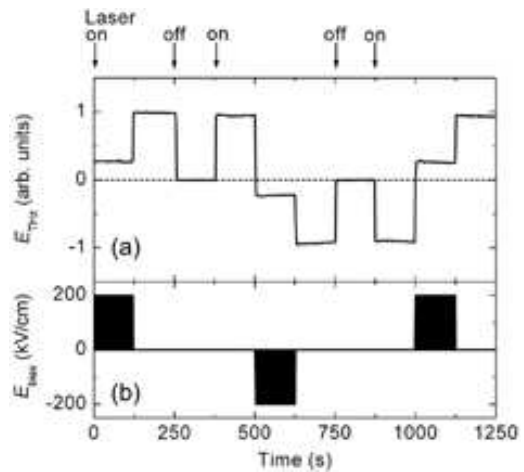}
\caption{Switching operation of BiFeO$_3$ photoconductive switch as a function of time probing with terahertz (THz) radiation. Temporal change of (a) THz amplitude $E_{\rm THz}$ and (b) the applied bias electric field $E_{\rm bias}$. The dashed line represents the zero-level line shown for clarity.}
\label{fig3}
\end{figure}

Such characteristics open a way to a variety of applications using THz radiation as a sensitive probe. The hysteretic feature demonstrates the potential of ferroelectric photoconductive devices to act as nonvolatile random access memories where the writing process will be performed by the application of $E_{\rm bias}$, and the reading by detecting THz radiation. This is accomplished by laser illumination and the ``0", ``1" information is determined by the signs of $E_{\rm THz}$, whether positive or negative. To explicit the switching operation, the temporal change of $E_{\rm THz}$ with varying $E_{\rm bias}$ is shown in Fig. \ref{fig3}(a). The application of positive $E_{\rm bias}$ of +200 kV/cm [Fig. \ref{fig3}(b)] writes down an information which is sustained even after the removal of $E_{\rm bias}$. THz radiation is indeed detected only when the laser is illuminated onto the film. In turn, the application of an opposite $E_{\rm bias}$ of $-200$ kV/cm erases the previous information and writes down a new information to the film. We confirmed that the written information was sustained for more than two weeks in the absence of $E_{\rm bias}$, and also that 12 hours of continuous illumination of laser pulses at zero-bias electric field did not give rise to any suppression of $E_{\rm THz}$ nor a change of its phase. Since the ``on" and ``off" operation of laser illumination does not induce a change of the $P_{\rm s}$ state, this process can be regarded as a nondestructive readout. Note that the rather high $E_{\rm bias}$ applied here can be reduced down to at least 50 or 60 kV/cm assuming from the coercive field estimated from Fig. \ref{fig2}(a).

\begin{figure}[bt]
\includegraphics[width=0.38\textwidth]{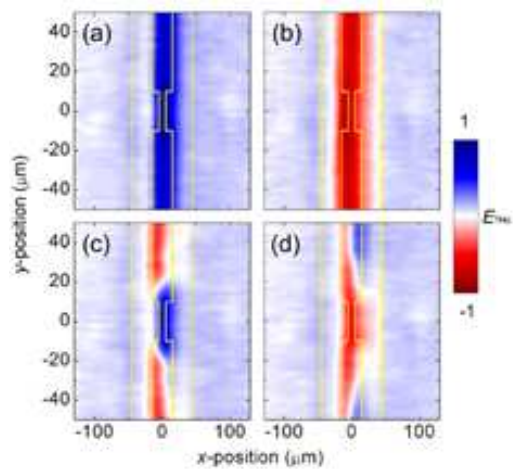}

\caption{(color) Visualization of 180$^\circ$ ferroelectric domain structure of the BiFeO$_3$ thin film probing with terahertz amplitude $E_{\rm THz}$. Domain structure after the application of bias electric field $E_{\rm bias}$ of (a) +200 kV/cm and (b) $-200$ kV/cm carried out with simultaneous laser illumination, and domain structure after the application of $E_{\rm bias}$ of (c) +200 kV/cm and (d) $-200$ kV/cm carried out without laser illumination. The geometry of the dipole-type Au electrodes is also shown by the yellow line in the respective images. Opposite 180$^\circ$ ferroelectric domains are distinguished by the blue and red colors based on the sign of $E_{\rm THz}$. The images showed no evolution with time as we repeated the scans of the same area for several days.}
\label{fig4}
\end{figure}

With the THz radiation exhibiting direct relationship with $P_{\rm s}$, scanning of the two-dimensional distribution of $E_{\rm THz}$ allows us to visualize the ferroelectric domain structures by distinguishing the orientation of the 180$^\circ$ domains from the signs of $E_{\rm THz}$. For demonstration, we applied this imaging technique to two different types of ferroelectric domain structures, which were realized by changing the poling conditions. In Fig. \ref{fig4}, we show four domain images of a section of the BiFeO$_3$ thin film at zero-bias electric field after the application of $\pm$200 kV/cm carried out with [Figs. \ref{fig4}(a) and \ref{fig4}(b)] and without [Figs. \ref{fig4}(c) and \ref{fig4}(d)] simultaneous laser illumination. The initial poling treatment of ``applying $E_{\rm bias}$ of $\pm$200 kV/cm with simultaneous laser illumination" is performed by once scanning the objective area at $\pm$200 kV/cm. This treatment was also carried out prior to the measurement in Figs. \ref{fig4}(c) and \ref{fig4}(d). Thus, the former state before the poling treatment of Fig. \ref{fig4}(c) [Fig. \ref{fig4}(d)] corresponds to the image shown in Fig. \ref{fig4}(b) [Fig. \ref{fig4}(a)]. Domains with opposite polarization states appear as blue and red areas depending on the sign of $E_{\rm THz}$. By comparing the images of each single pairs [between Figs. \ref{fig4}(a) and \ref{fig4}(b), as well as Figs. \ref{fig4}(c) and \ref{fig4}(d)], one can see that only the domains between the electrodes has changed their states by the application of opposite $E_{\rm bias}$, while the other areas are independent of $E_{\rm bias}$ and remain unchanged. The different poling treatment, whether $E_{\rm bias}$ was applied with or without laser illumination, gave rise to an interesting feature as demonstrated by the comparison of the pair of Figs. \ref{fig4}(a) and \ref{fig4}(b) by the pair of Figs. \ref{fig4}(c) and \ref{fig4}(d). In Figs. \ref{fig4}(a) and \ref{fig4}(b), the entire domains between the electrodes have managed to reverse their phase by $\pi$. Contrary, in Figs. \ref{fig4}(c) and \ref{fig4}(d), only the area near the dipole gap has reversed its domains where maximum electric field is applied due to geometric effect, while the other area between the two strip lines has not, realizing the coexistence of two opposite 180$^\circ$ domains aligned side by side between the electrodes. The origin of the two different states can simply be understood by the term ``photoassisted $P_{\rm s}$ switching" previously reported in BaTiO$_3$ \cite{16} and Pb(Zr,Ti)O$_3$ \cite{17} thin films where the combination of ultraviolet light exposure with applying $E_{\rm bias}$ showed superior switching ability than the application of $E_{\rm bias}$ alone. It is also noteworthy here that we observed THz radiation even in the areas apart from the electrodes where the domains had not been artificially aligned. However, the intensity of $E_{\rm THz}$ in these areas is considerably weak presumably because the domains are not aligned neatly as in the areas between the electrodes. These imaging results indicate that THz radiation occurs from individual ferroelectric domains and the intensity of $E_{\rm THz}$ is determined by the average of the electric dipole moments of which the laser spot covers. Since the spatial resolution of this imaging technique is limited by the diameter of the focused laser spot, we assume that a more detailed domain structure can be observed by focusing the laser spot down to a submicrometer scale. This guarantees the potential of the present technique to compete evenly with the other fascinating ferroelectric domain imaging tools such as atomic force microscopy \cite{18} and x-ray microdiffraction \cite{19}.

In conclusion, we have demonstrated a novel THz radiation functionality of a multiferroic BiFeO$_3$ thin film triggered via ultrafast modulation of $P_{\rm s}$ upon carrier excitation by the illumination of femtosecond laser pulses. By utilizing the hysteretic THz radiation characteristic where $E_{\rm THz}$ directly reflect the $P_{\rm s}$ state, we have also presented potential approaches to nondestructive readout in nonvolatile ferroelectric memory and 180$^\circ$ ferroelectric domain imaging system. Since BiFeO$_3$ belongs to a class of multiferroics, measurement under magnetic field is promising, for it may give rise to an additional degree of freedom in the THz radiation feature also providing an additional functionality in device designing.

The authors thank M. Suzuki, R. Inoue, and K. Kotani for fruitful discussions. This work was supported by the Strategic Information and Communications Research and Development Promotion Fund of Ministry of Internal Affairs and Communications (MIC). K. T. acknowledges the financial support from JSPS.

\end{document}